\documentclass[acmsmall]{acmart}
\usepackage{tabularx}
\AtBeginDocument{%
  \providecommand\BibTeX{{%
    \normalfont B\kern-0.5em{\scshape i\kern-0.25em b}\kern-0.8em\TeX}}}

\begin{document}

\title{Languaging Ethics in Technology Practice}


\author{Colin M. Gray}
\email{gray42@purdue.edu}
\affiliation{%
  \institution{Purdue University}
  \streetaddress{401 N Grant Street}
  \city{West Lafayette}
  \state{Indiana}
  \postcode{47907}
  \country{USA}
}

\author{Shruthi Sai Chivukula}
\email{schivuku@iu.edu}
\affiliation{%
  \institution{Indiana University}
  \streetaddress{901 E 10th St Informatics West}
  \city{Bloomington}
  \state{Indiana}
  \country{USA}
}

\author{Janna Johns}
\email{johns20@purdue.edu}
\author{Matthew Will}
\email{will10@purdue.edu}
\author{Ikechukwu Obi}
\email{obii@purdue.edu}
\author{Ziqing Li}
\email{li3242@purdue.edu}
\affiliation{%
  \institution{\newline Purdue University}
  \streetaddress{401 N Grant Street}
  \city{West Lafayette}
  \state{Indiana}
  \postcode{47907}
  \country{USA}
}

\renewcommand{\shortauthors}{Author et al.}

\begin{abstract}
Ethics as embodied by technology practitioners resists simple definition---particularly as it relates to the interplay of identity, organizational, and professional complexity. In this paper we use the linguistic notion of \textit{languaging} as an analytic lens to describe how technology and design practitioners negotiate their conception of ethics as they reflect upon their everyday work. We engaged twelve practitioners in individual co-creation workshops, encouraging them to reflect on their ethical role in their everyday work through a series of generative and evaluative activities. We analyzed these data to identify how each practitioner reasoned about ethics through language and artifacts, finding that practitioners used a range of rhetorical tropes to describe their ethical commitments and beliefs in ways that were complex and sometimes contradictory. Across three cases, we describe how ethics was negotiated through language across three key zones of ecological emergence: the practitioner's ``core'' beliefs about ethics, internal and external ecological elements that shaped or mediated these core beliefs, and the ultimate boundaries they reported refusing to cross. Building on these findings, we describe how the languaging of ethics reveals opportunities to definitionally and practically engage with ethics in technology ethics research, practice, and education.

\end{abstract}

\ccsdesc[500]{Human-centered computing~Empirical studies in HCI}
\ccsdesc[500]{Social and professional topics~Computing industry}
\ccsdesc[500]{Social and professional topics~Codes of ethics}

\keywords{ethics, technology practice, languaging, rhetoric, co-creation}

\maketitle

{\color{red} \textbf{Draft: April 12, 2023}}

\section{Introduction}

The ethical dimensions of technology practice are increasingly central in Human-Computer Interaction (HCI) and design scholarship, impacting both public policy and the role(s) that practitioners are able to take on in considering social impact and their responsibility in relation to this impact~\cite{shilton2018values,Lazar2015-lh}. HCI and Science and Technology Studies (STS) scholars have addressed ethics through a broad range of framings, frequently articulated in relation to dominant paradigms of ethics (e.g., consequentialist, deontological, virtue, care;~\cite{Becker2013-wz,shilton2018values}), critically-oriented epistemologies (e.g., feminism~\cite{Bardzell2010-pw,Chivukula2020-zl}, social justice~\cite{Dombrowski2016-go,Costanza-Chock2020-mk}, critical race theory~\cite{Ogbonnaya-Ogburu2020-gq}), and drivers for organizational and professional responsibility~\cite{Papanek1971-fv,Monteiro2019-tj,Van_Wynsberghe2014-wf}. Dominant strands of research have proposed methodologies for ethics-focused engagement \cite{friedman2013value,Flanagan2014-hf}, identified engagement strategies to raise matters of ethical concern in everyday work \cite{Wong2021-pv,shilton2013values,Lindberg2021-hi}, and revealed aspects of design complexity that mediate the potential for ethical engagement \cite{gray2019ethical,Watkins2020-zr,chivukula2021identity}.

However, the notion of ethics is not a simple one. Researchers have variously centered their engagement around pragmatic links to moral philosophy~\cite{Friedman2019-zg,frauenberger2017action}, the imperative of disciplines to take responsibility for social impact \cite{Costanza-Chock2020-mk,Gotterbarn2017-nw}, professional codes of ethics~\cite{Gotterbarn2017-nw,Buwert2018-uw}, organizational complexity and regulatory control~\cite{Gray2021-zf,Perezts2015-tz,Wong2023-ae}, and considerations of the practitioner as the ``guarantor of design'' \cite{Nelson2012-ov}. What ``ethics'' means in each of these contexts, whose ethics they are, and who is considered as a stakeholder can vary widely based on these potential anchors or areas of focus. 
To understand this diversity of ethics and how it is framed in practitioner descriptions of their felt complexity, we sought to identify a more precise vocabulary through which researchers might productively engage in research that relates to ethics and technology and design practice---not to provide absolute conclusions on which definitions of ethics are appropriate or useful, but rather to identify how language as a world-building activity~\cite{Krippendorff1995-bq,Demuro2021-io} that is undertaken by all humans can reveal hidden assumptions and tacit conceptions of ethics on an individual level.

In this paper, we describe how a range of technology practitioners revealed aspects of their ethical standpoint through the linguistic frame of \textit{languaging}. We analyzed a set of 90--120 minute co-creation sessions that we conducted with twelve individual practitioners representing multiple professional roles. Through a linguistically-focused analytic frame, we identified instances where practitioners reasoned about ethics in language and through co-creation artifacts, indicating ethical aspects of their own professional role and relationships between this role and their ecological setting of practice. We evaluated these instances using an existing set of rhetorical tropes, elucidating the ways in which practitioners used \textit{synecdoche, metonymy, antithesis,} and \textit{amplification} to encompass and triangulate their notions of ethics. By documenting practitioners' use of this range of tropes, we reveal the complex emergence and diversity of ethics as a construct in relation to technology practice. We then more fully explore this diversity in an ecological sense across three case studies, where we identify the ``\textit{core}'' metaphors used by practitioners to frame ethics, the \textit{internal emergence} of this core in relation to the practitioner's identity and professional role, the \textit{ecological emergence} of the core and identity in everyday work, and the ultimate \textit{ethical boundaries} that practitioners were unwilling to cross. This analysis further reveals the complexity of ethics as a construct in technology practice and demonstrates the utility of linguistic analysis to aid in deconstructing the internal coherence of various conceptions of ethics by framing this coherence as languaging.

The contribution of this work is three-fold. First, we identify the ways in which rhetorical tropes are used by technology practitioners to tangentially and directly define ethics, providing a robust analytic frame for researchers, practitioners, and educators to reflect upon their ethical commitments in ways that link subjective position to broader ``worlds'' of ethical engagement. Second, we reveal the complex interplay of ethics as a concept across individual and ecological dimensions, extending past research on ethical complexity and sensitivity and revealing new spaces for intervention, reflection, and reconceptualization of technology ethics. Third, we offer a linguistically-focused analytic lens to deconstruct and describe ethical complexity in technology practice, providing researchers and technology practitioners with a pathway towards building mutual understanding regarding different conceptions of ethics.  

\section{Related Work}

\subsection{Ethical Complexity in Technology Practice}
Numerous HCI and STS scholars have studied the ethical complexities of technology practice through a range of different lenses~\cite{lindberg2020cultivating, Lindberg2021-hi, shilton2013values, friedman2013value, waycott2016ethical, molich2001ethics,Wong2021-pv}. For example, Shilton~\cite{shilton2013values} and Friedman et al.~\cite{friedman2013value} examined how design practices influence the social values that are embedded in products and how designers account for those values when developing systems. Lindberg et al. ~\cite{lindberg2020cultivating} engaged with practitioners to learn about their understanding of ethics and to examine the challenges they face infusing ethics into their professional practice. Their end goal was to develop a set of actionable questions that will help advance `ethos' in HCI practice. Wong ~\cite{Wong2021-pv} studied the tactics user experience professionals employ to infuse values into their practice, surfacing how those tactics allow them to induce value-oriented outcomes in their organizations. Shilton ~\cite{shilton2018values} traced and examined historical attempts to incorporate values into HCI and technology practice. And Boyd and Shilton ~\cite{Boyd2021-sv} revealed how ethics is operationalized in technology design teams, including moves that include recognition, particularization, and judgment. Other scholars ~\cite{waycott2016ethical, molich2001ethics, frauenberger2017action} have also studied and discussed the dilemmas practitioners face navigating and translating ethical contexts into their everyday practice. Relevant here, is that these researchers all acknowledge the complexity of introducing ethics into technology practices and the need to provide support to practitioners to enable ethical awareness and the potential for action.

Although discussions on the importance of ethics to technology practice are ongoing~\cite{shilton2018values}, practitioners and researchers still grapple with concretely describing the interplay of ethics and practice ~\cite{cairns2003diversity,gray2019ethical}. Munteanu et al. ~\cite{munteanu2015situational} studied how ethical frameworks typically designed to guide practitioners are difficult to follow in real-life settings because they often do not capture all the interactions of technology practitioners within their organization. Gray and Chivukula ~\cite{gray2019ethical} examined the decision-making process of design practitioners, highlighting how organizational, individual, and external descriptions of ethics influences the way practitioners interpret and navigate ethical contexts in professional practice. They describe the mediation that emerges among these elements as a felt \textit{ethical design complexity} that practitioners face in technology practice. They later built on this work to describe the individual identity claims that practitioners come in with ~\cite{chivukula2021identity} in relation to their profession, their organizational role, and how their personal beliefs impacted how they were able to engage in those conversations and ethical contexts. 
These studies and numerous others (e.g.,~\cite{brown2016five, Steen2015-qw,shilton2018values, mcmillan2019against, gray2018dark, mathur2019dark}) highlight the need to further engage with practitioners to explore and describe how they reason about ethics in their everyday practice in ways that are situated, contingent, and complex.

\subsection{Languaging and Rhetoric}
Rhetoric, a conceptual framing of communication invented by ancient philosophers, has long been used by scholars as a means by which language can be deconstructed, interpreted, and shaped. and construct complex language. While rhetoric has inspired a range of disciplinary perspectives that are common in HCI scholarship, the use of rhetoric as an explicit analytic vocabulary to investigate designer practices is somewhat more rare. In the past decade, design and HCI scholars such as Sosa-Tzec~\cite{Sosa-Tzec2015-vu,Sosa-Tzec2015-jp,Sosa-Tzec2014-nx} and Halstr{\o}m~\cite{Halstrom2016-ir,Halstrom2017-ny}, have used rhetoric as a lens to explore different aspects of visual communication and engagement in design activity, including persuasion, meaning, argumentation, connection with oral communication, and generation of concepts. Halstr{\o}m \cite{Halstrom2016-ir} used the language of rhetoric to describe how designers engage in a ``celebration of values'' as they frame and navigate the design situation, leading to a positioning of design outcomes ``[\ldots] not just [as] a solution that persuades others to buy it, but one that celebrates the values we intend.'' In a related paper, Halstr{\o}m~\cite{Halstrom2017-ny} described how rhetoric can be used to support design argumentation, such as how components of amplification can ``be used as a rhetorical framework for designers to explore various ways of amplifying `the claim' they are making in the form of design artifacts.'' Sosa-Tzec has used rhetorically-focused vocabulary to support investigations of how designers inscribe purposive visual and interactive elements into designed outcomes with the intent of persuasion and appeal, such as in his investigation of \textit{Gaza Everywhere}~\cite{Sosa-Tzec2015-jp}. In Sosa-Tzec's \textit{Rhetorical Handbook}~\cite{Sosa-Tzec2014-nx}, he translated many common elements of rhetoric vocabulary from a visual communication to interaction design/UX context, demonstrating both the range of vocabulary available and its use in describing complex attributes of designed experiences. We leverage his description of rhetorical tropes such as synecdoche, metaphor, metonymy, amplification, and antithesis as a means to examine and communicate design complexity through language, with a focus not on visual artifacts that result from design processes, but rather accounts from technology practitioners in relation to the complexity through which design work happens. 

We use this perspective towards rhetoric, relying on the ways that language is used to construct our social world. This idea is not new---but rather builds upon over 50 years of inquiry that relate to linguistics and action. Prominent examples of this approach include J. L. Austin's Speech Act Theory~\cite{Austin1975-dv}, where words are shown to have not just descriptive, but also performative and social power (e.g., illocutionary speech acts that bring with them performative force, such as ``You're fired'' or ``I pronounce you married for life''). The work of Wittgenstein~\cite{Wittgenstein2013-lc} also connects words and their meaning to their presence in particular ``language games,'' whereby words that can have multiple meanings take on a more specific meaning, and with that meaning shape the realities that we live within and through. We do not intend to exhaust the literature relating to this framing of communication, but rather take on this constructive view of language as a tool for saying, doing, interpreting, and acting. Through these activities, we construct our social world, form normative notions of concepts, and establish our own identity in relation to others through language.\footnote{There is a strong overlap between these broader linguistic tools and those that have been operationalized in contemporary discourse analysis, such as through the foundational work of Gee~\cite{Gee2014-iv}. Because we standardize our analysis on tropes, we do not employ any of Gee's discourse analysis tools directly, but many links could be drawn between these complementary methodologies.} To do so, we leverage one particular perspective on rhetoric and linguistics, focusing on how we \textit{language} our world into reality---whereby the language we use (or don't use) constructs and brings into being our sense of reality and our possibility for action within that reality (cf.,~\cite{Demuro2021-io,Krippendorff1995-bq}. As an example of the power of languaging in a design context, Jony Ive (then Chief Design Officer at Apple) noted how even simple word choice can open up or shut down the rhetorical space: ``\textit{If we’re thinking of lunchbox, we’d be really careful about not having the word `box' already give you bunch of ideas that could be quite narrow. You think of a box being a square, and like a cube. And so we’re quite careful with the words we use, because those can determine the path that you go down}'' \cite{Cheng2013-pl}. Similarly, language in design practice pre-frames the situation, raising certain matters of concern to the foreground and moving others to the background, expressing an ontological stance that is simultaneously normative and subjective (cf., \cite{Willis2006-qb}). In this paper, we take on this frame of languaging to more fully describe how word choices in context reveal broader conceptual understandings that technologists have about complex concepts---in our case, ethics in technology practice.

\section{Method}
We used a co-creation approach \cite{Sanders2008-nm} to engage practitioners from different professional roles in communicating about ethics in their everyday work. We designed three broad co-creation activities as part of a larger study to encourage technology and design practitioners to reflect on their ethical awareness, responsibility, and potential for action. The analytic focus of this paper is not on the co-creation activities themselves, but rather the conversations that resulted from practitioners' engagement with the co-creation material as they reflected on their ethical role in their organization. After extensive reflexive engagement with the co-creation session transcripts \cite{Braun2019-hc}, we focused on \textit{languaging} as our primary analytic focus, since it allowed us to describe the normative assumptions that were present in the transcripts. In this paper, we seek to answer the following research questions:
\begin{enumerate}
    \item How was ethics languaged by technology and design practitioners?

    \item How can multiple elements of languaging be used to reveal the dynamic negotiation of practitioners' ethical standpoints? 
\end{enumerate}

To answer these research questions, we conducted a reflexive thematic analysis \cite{Braun2019-hc}, inspired by Sosa-Tzec's \textit{Rhetorical Handbook} \cite{Sosa-Tzec2014-nx}, which had previously translated aspects of visual rhetoric for an HCI audience. We used the tropes in this source, alongside broader characterizations of tropes in the rhetoric literature, to identify the different \textit{rhetorical tools} that practitioners used to language ethics through a top-down thematic analysis. We then augmented this top-down analysis with three cases that describe how different practitioners used multiple rhetorical tools in combination to make sense of their own ethical role, focusing on illustrating the \textit{interactional qualities} of the rhetorical tropes in relation to each other and in relation to other ecological factors that shaped the practitioner's engagement with ethics. In the following sections, we describe the co-creation activities, the sampling and data collection procedures, and the analysis process.  

\subsection{Co-Creation Activities}
We used a co-creation methodology \cite{Sanders2008-nm} to design three activities. The primary goal of these activities was to engage and support practitioners in communicating about their felt ethical concerns both in their primary professional role and as they interacted with practitioners from other roles.  
The first activity engaged practitioners in 
visually mapping different people, factors, and structures 
that impacted their ethical engagement, with the goal of identifying the kinds of support they desired to increase their ethical engagement. 
The second activity engaged practitioners in considering a range of ``ethical dilemmas'' captured in a previous interview study, with the goal of encouraging the practitioners 
to reflect, elaborate, and speculate about their 
ethical commitments, decision making, and complexity in relation to their everyday work. 
The third activity engaged practitioners in evaluating an existing ethics-focused method through the application of a set of method heuristic tags, with the goal of identifying how and whether the method resonated with their ecological setting. Our analytic focus in this paper is on what these activities helped participants elicit in downstream conversation rather than through the activities themselves. 

\subsection{Sampling of Participants}
Participants were recruited through a screener posted on a range of social media sites, including Twitter, LinkedIn, Facebook, and Reddit. We also sent the screener to participants from a large interview study that we had previously conducted on technology ethics in organizations. We sought to sample a range of professional roles, while also seeking diversity in organization type, years of experience, and gender.
Participants are referred to through a unique identifier in the format P0n, where \textit{n=1--12} represents the number assigned to the twelve practitioners. Participant roles included: UX designers (P01, P07, P10), UX researchers (P04, P11), Software Engineers (P06, P09, P12). Product Managers (P02, P05, P08), and a Data Scientist (P03). Participants worked in organization types such as Enterprise (B2B and B2c), Agency, or Consultancy and had professional experience ranging from 1--8 years (Average=3.7 years).

\subsection{Data Collection}
We engaged each practitioner in two of the three co-creation activities for 90--120 minutes via Zoom, with one researcher facilitating each session. The co-creation activities were designed and conducted on a digital collaborative digital whiteboarding tool called Miro\footnote{https://miro.com/}. The engagement between the practitioner and the facilitator was audio and video recorded for analysis. Practitioners consented to this study, which was approved by our Institutional Review Board (IRB).

In total, we facilitated approximately 23 hours (1385.3 minutes) of co-creation sessions with an average engagement of 115 minutes per session across the twelve practitioners. All sessions were video recorded and transcribed, and these transcripts were the main focus of analysis for this research paper, which captured the conversations with the practitioners including interactions during the activities, asking of clarifying questions, ideation of new possibilities through the activities, and debriefing conversations at the end of the engagement.

\subsection{Data Analysis}
We conducted analysis in four phases: 1) building familiarization with the collected data, 2) identifying vignettes in the transcripts, 3) coding the identified vignettes, and 4) building and visualizing individual cases to elaborate and extend the thematic coding. The research team for data analysis consisted of a principal investigator and five trained graduate and undergraduate researchers who had experience with qualitative research analysis through academic coursework or other research projects. Throughout the analysis process, we included rounds of debriefing and member checking which allowed us to be intentionally reflective and reflexive. We detail the four phases, our researcher positionality, and exclusion criteria in the sections below. 

\subsubsection{Familiarization} 
We began by familiarizing ourselves with the data collected. Data familiarization was conducted using a qualitative analysis tool and consisted of listening to and reading the entirety of each transcript and in an exploratory manner, using first-cycle open coding~\cite{Saldana2015-ey} to identify salient aspects of the stories shared by practitioners. Through these codes, we identified multiple aspects of ethics shared by the practitioners, including their work ecologies, decision-making processes, supports and barriers for ethical decision making, interactions within their ecology, personal ethics, and disciplinary notions of ethics. We then debriefed as a research team, converging on a how practitioners used language relating to ethics as they interacted with the co-creation activities. We had chosen not to define ethics to the practitioners during the workshops and framed the activities for practitioners with the intention for them to reflect on ``their ethical responsibility, awareness, and action'' as they felt it fit their context. Given the generative nature of the material provided, we identified potential in how practitioners' use of language indirectly revealed their definitions and felt complexity of ethics, which framed the focus of this paper.

\subsubsection {Vignette Identification}
Building on our first-cycle coding, we identified ``vignettes'' which included instances where participants described or referenced ethics \textit{in their own words,} either directly or through low inference speech acts. For example, while participant P04 was describing her work for a client, creating a medical billing processor, she stated: ``\textit{I had been pushing for accessibility design, knowing this was going to a general population}.'' We coded this instance as an eligible vignette for further analysis because P04 expressed a relational stance towards ethics using the language of ``accessibility'' experienced by a ``general population.'' We excluded instances where participant's languaging was overtly influenced by the co-creation activity, material, and vocabulary. For instance, the second activity provided the participants with a list of ethical dilemmas which has been preframed as relating to ethics. To finalize the vignette identification, we engaged in multiple rounds of debriefing and reflexive engagement as a group of researchers, where we asked questions such as: ``Are they saying this is what ethics means to them?'' or ``Are the words being used expressing underlying values or ethical commitments?'' Each transcript was assigned to two researchers---a primary and a secondary coder. The two coders member-checked each other's vignettes to add, contest, and identify any discrepancies in the vignette identification, resulting in full agreement. In total, we coded 394 vignettes across the twelve transcripts. After we decided to focus our analysis explicitly on the languaging of ethics, we excluded 282 vignettes that focused only on the origin of the practitioner's understanding of ethics. To answer our first research question, we decided to focus on the remaining 112 vignettes for further analysis, while using the other vignettes to inform our understanding of the infrastructure of languaging through case analysis to support answering our second research question. 

\subsubsection{Reflexive Thematic Coding}
To describe how practitioners languaged ethics, we used a set of existing rhetorical tools to code the identified subset of vignettes. We were inspired by a set of rhetorical tropes defined in Sosa-Tzec's Rhetorical Handbook \cite{Sosa-Tzec2014-nx}, which---while not focused explicitly on ethics---was one of the first direct translations of rhetorical tools into HCI scholarship. Sosa-Tzec describes a \textit{trope} as a rhetorical tool that is used to ``\textit{alter the normal reference of the elements}'' in place (cited in \cite{Ehses1988-vx}). The six tropes outlined in the handbook are consistent with broader sets of tropes referenced in the rhetoric literature, including: \textit{synecdoche, antithesis, metonymy, amplification, metaphor,} and\textit{ personification}. As a group, we coded all the vignettes using these six tropes non-exclusively. This analysis revealed that practitioners most frequently used synecdoche as a rhetorical tool to talk about ethics; very rarely used metaphors, such as a participant comparing web-scraping to a ``gray area'' within ethics; and never used personification. Therefore, we decided to exclude metaphors and personification from our analysis. The definitions of these tropes and our operationalization to focus on the languaging of ethics are listed in Table~\ref{tab:tropes}. 

\begin{table*}[!htb]
\caption{Rhetorical Tropes and Definitions}
\label{tab:tropes}
\begin{tabularx}{\textwidth}{ p{8em} p{13em} X }
\toprule
\textbf{Rhetorical Trope} & \textbf{Definition from Sosa-Tzec's Rhetorical Handbook \cite{Sosa-Tzec2014-nx}} & \textbf{Operationalized as...} \\
\midrule
Synecoche & Using a part of an object to represent a whole. & Describing ethics by \textit{referring} to their conceptual understanding of ethics as a part of a whole or vice-versa. \\
Metonymy & Representing one term with another which is close to it in time, space, or causation. & Describing ethics by \textit{replacing} ethics with a related term or concept. \\
Antithesis & Contrasting two opposing objects or ideas. & Describing ethics by \textit{contrasting} two or more different ideas relating to ethics, often framing what qualifies as being ethical or \textit{not} ethical. \\
Amplification & Discussing in detail the parts of an object or argument. & Describing ethics by \textit{foregrounding} their ethical commitment, awareness, action, and responsibility through detailed descriptions or examples.\\
\bottomrule
\end{tabularx}
\end{table*}

In our findings, we report the four tropes that were most frequently used by our participants. While analyzing the vignettes across this set of tropes, we also identified interactional or dynamic qualities of the tropes (e.g., amplification being used to detail synecdoche) and the influence of ecological, disciplinary, personal, and experiential factors of the practitioner on the tropes. This prompted us to activate the participants' use of tropes in more detail through a case study analysis.
 
\subsubsection{Case Identification and Analysis}
In this section, we present our second form of analysis drawing from case study methodology \cite{Yin2009-vs}, where the participant was the unit of analysis. After completing our trope analysis, we identified various interactional qualities of the rhetorical tropes and differences among the participants in how they were languaging ethics based on their professional role, personal factors, and ecological factors. Across the twelve participants, we identified three distinct cases to exemplify the range of interactions of various tropes used by the practitioners with a focus on how languaging functioned across their ecological, disciplinary, societal, and personal operationalization of ethics. We chose P02, P08, and P11 based on multiple criteria. Demographically, they are from different professional roles: P02 is a product manager (PM), P08 is a product manager who has recently transitioned from being a software engineer, and P11 is a UX researcher. These three cases also represented differing years of experience across different industry types, and included a variety of languaging of ethics illustrated through different combinations of rhetorical tropes. 

To illustrate the functioning of various tropes, we reflexively designed a visual schema and used it to map each of the three practitioner cases. The goal for this schema was not to reduce complexity or assert the existence of only certain interactional types, but rather---in building on Nelson and Stolterman’s definition of schema---``\textit{represent holistic concepts, ideas, and fundamental knowledge in visual form [\ldots with the goal of] expand[ing] and complement[ing] the text in revealing or reflecting deeper understandings of design}'' \cite{Nelson2012-ov}. The schema consists of four concentric zones or \textit{layers}, as shown in Figure \ref{fig:layerskey}: 

\begin{figure}
    \centering
    \includegraphics[width=0.5\textwidth]{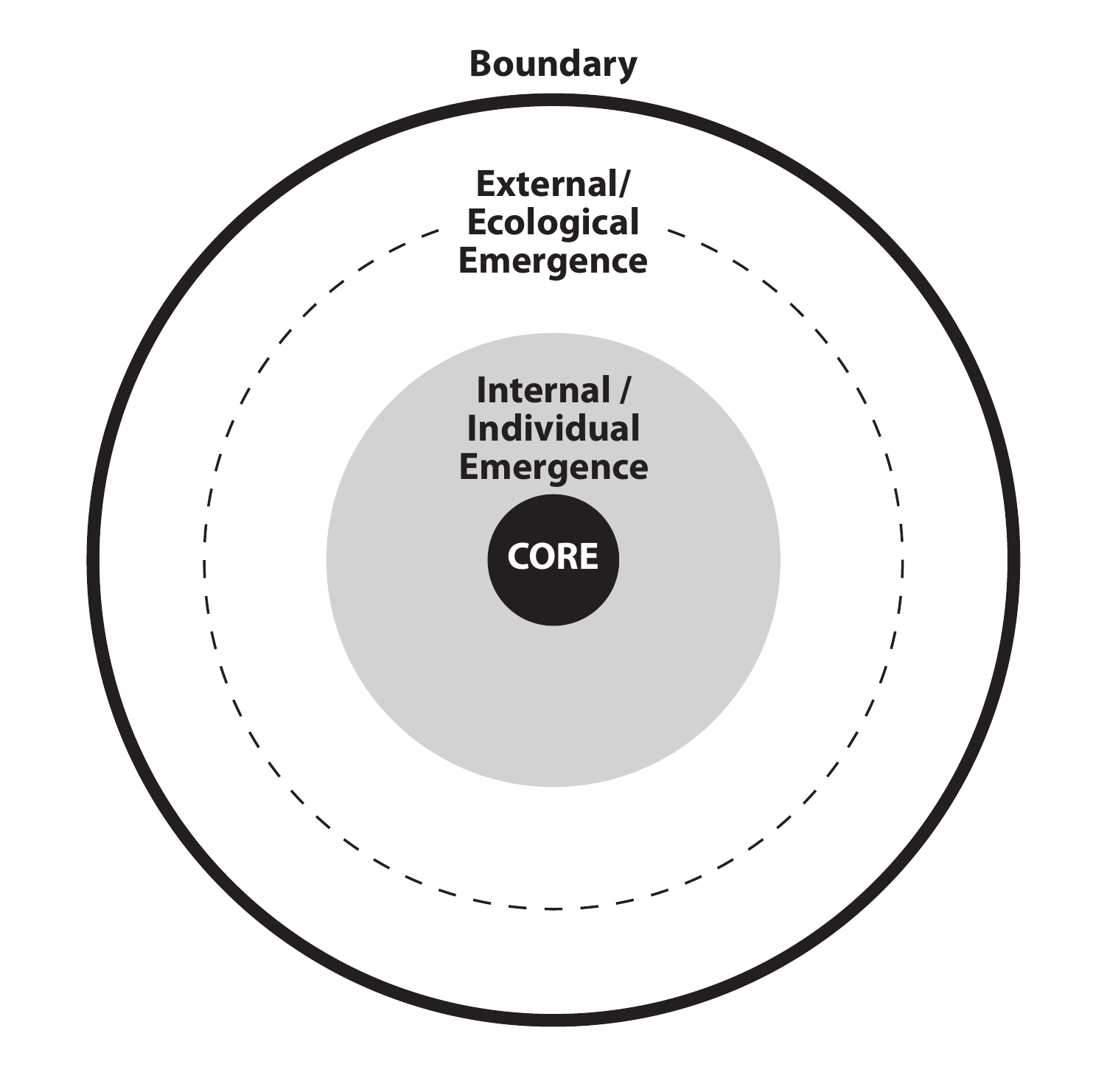}
    \caption{Layers of Languaging Schema}
    \label{fig:layerskey}
\end{figure}

\begin{itemize}
\item \textbf{Core}: The center circle is the \textit{core}, representing the key metaphor that describes the participant's foundation for ethical decision-making, commitment, action, and awareness. The core parallels prior work on identity claims of technology practitioners that inform subsequent action~\cite{chivukula2021identity}.
\item \textbf{Internal/Individual Emergence}: The core is surrounded by a zone wherein practitioners \textit{individually and internally} build on, inform, or question their core through language. This zone of individual emergence parallels the interaction and prioritization of differing identity claims~\cite{chivukula2021identity}.
\item \textbf{External/Ecological Emergence}: This zone is marked by a permeable ring (represented by a dotted line) that accounts for a practitioner's actions and views that they can or cannot perform as defined or shaped by their current ecology, interactions with other professional roles, or other external societal factors. This zone of ecological emergence parallels prior work on ecological engagement, including value levers~\cite{shilton2013values}, ethical mediation~\cite{gray2019ethical}, and soft resistance~\cite{Wong2021-pv}.
\item \textbf{Boundary}: The ultimate boundary defined ``limit'' or ``edge'' cases (represented by a thick line) that a practitioner explicitly states will not be crossed due to a combination of their core, professional role, or ecological factors. 
\end{itemize}

The \textit{core} and the \textit{internal/individual emergence} form the practitioner's individual perspective in relation to ethics, which is then performed and shaped in relation to their surrounding work environment as represented by the \textit{external/ecological emergence} layer. Relationally, languaging is used both to form relationships among the former three layers and to identify the ultimate \textit{boundary} which when crossed would be not ethical or consistent with their core. 

As a research team, we used both the direct and infrastructural vignettes to identify how each participant case leveraged languaging across these four layers, with the goal of describing the internal coherence of each practitioner's ethical commitment, action, and responsibility. Through this process, we identified ``rhythms'' of languaging, where patterns reflected interactions among the various layers, or the use of combinations of rhetorical tropes to strengthen, contest, contrast, or shape their ethical standpoint through language. We detail the resulting cases using this schema in Section \ref{cases}.

\section{Languaging Structures} \label{tropes}
In this section, we answer RQ\#1 by describing how ethics was languaged using Synecdoche, Metonymy, Antithesis, and Amplification (Table \ref{tab:tropes}). We provide definitions of the tropes and identify the different ways practitioners used these tropes to language ethics. 

\subsection{Synecdoche}
Synecdoche was the most frequently used rhetorical tool used by practitioners to describe ethics,  referring to their conceptual understanding of ethics as a part of a whole or a whole in relation to existing parts. Participants used synecdoche in three ways: 1) \textit{Part to Whole} relationships, where practitioners communicated their ethical alignment or commitment by using a particular value or lens (part) of ethics (whole), 2) \textit{Cluster to Whole} relationships where practitioners gathered a cluster of ethical commitments or values to communicate about ethics (whole), and 3) \textit{Lenses} described practitioners' use of a particular perspective on their ethical action that instantiated the referred parts, clusters, or whole of ethics. 

\subsubsection{Part to Whole Relationship}
Practitioners used synecdoche to describe a part-whole relationship, frequently using \textit{parts} of ethics such as transparency, bias, data privacy, accessibility, and legality to refer to \textit{their whole} of ethics. For example, practitioners commonly referred to \textit{accessibility} as their focus of ethical action, while also referring to this term in different ways with different forms of operationalization. Participant P10 described ``accessibility and all that stuff'' as her focus while designing websites, pointing towards a broader conception of ethical engagement that could be referenced or defined through the lens of accessibility. P04, P07, and P11 also used accessibility as a primary frame for ethics, with P04 mentioning ``\textit{I was leveraging some of the accessibility standards to say that we could be legally responsible.}'' In a similar ``checklist'' approach to ethics through the lens of accessibility, P07 provided an example about their design of products that are ``accessibility complaint'' as required by WCAG guidelines; and P11 expressed their desire to make ``everything [e.g., digital product features] completely accessible for everyone because it's the right thing to do.'' In all the above instances, the term ``accessibility'' was used to language their ethical commitment in their design work---activating a broad set of work practices and set of conclusions about the ethical inscription process with particular limitations and beneficiaries. The various ways practitioners translated accessibility illustrates the potentially expandable nature of that part-whole relationship, where accessibility stands in for many upstream and downstream decisions. Transparency was another common word used by participants to language their ethics as a whole. For example, P11 described an insurance project he was working on: ``\textit{people don't trust anything to do with insurance, so we---you know---in terms of doing what is best for the customer and try to promote transparency.}'' Here, P11 indicated his goal of promoting transparency, where transparency stands in for a broader---yet unstated---ethical whole. P01 also asserted privacy as part of her ethical standpoint, noting: ``\textit{policies definitely affect the product, especially like privacy and things like that, but that's great.}'' In all the above examples, the ``parts'' used by the practitioners to languaging ethics are likely to have been influenced by their expectations of professional responsibility, ecology, product domain, and/or their personal values. However, it is interesting to note the frequent use of these ``parts'' to focus on one aspect of their ethical commitment in their work. 

\subsubsection{Cluster to Whole Relationship}
Practitioners used synecdoche through cluster-whole relationships when they referred to a group of values, a conceptual frame, or domain related to ethics (cluster) to represent the whole of ethics. Two common clusters used by practitioners were \textit{responsibility} and \textit{research ethics}. 

Practitioners used ``responsibility'' as a conceptual cluster to engage with ethics more broadly, using the breadth of this term to describe a range of component issues. P04 illustrates this conceptual clustering as an overarching umbrella for ethical engagement, stating: ``I feel this like responsibility to do my due diligence to make sure that we're not causing harm and we're doing our clients justice as well. Like we're delivering the right product in the most ethical way.'' The pieces within this example which contribute to P04's responsibility cluster include ``not causing harm,'' ``doing our clients justice,'' and ``delivering the right product.'' These three parts are separate, yet link to the broader conceptual whole of ``responsibility,'' languaging what responsibility means for the participants' ethical engagement and delineating a potential space and set of activities for ethical engagement. Other participants, such as P08, also used this clustering technique to include both what the cluster did include and did not include (cf., antithesis): ``My only job is to basically write the code that achieves those features, that implements those solutions. So I'm not worried about what are the real life applications, because it's, unless it's too dangerous---As a software engineer, I mean, it's not my responsibility.'' This cluster of responsibility as a software engineer includes ``writing code'' and ``not causing danger in doing so,'' while it is others' responsibility to consider the ``real life applications.'' P12 also identified their own responsibility in relation to others, stating: ``\textit{I would say the concentration should be on dividing the responsibility, but I think it doesn't mean that we should ignore some thing that might not be our entire responsibility. But when its the general goal, it should be taken into consideration, even though it might not personally be your responsibility.}'' In this example, P12 viewed responsibility as a cluster that should be divided, however certain circumstances required focus on the cluster of responsibility in relation to ethics as a whole. 

Practitioners used \textit{research ethics} as a cluster to indicate a broad range of actions such as consenting procedures, representative sampling strategies, and validity of research procedures. They used the cluster of research ethics to point towards broader, yet often unstated, ethical commitments. P04 sought to include the right participants for a research study addressing the elderly population; although she received no budget for the research, she valued having the right voice behind her work and used the sampling of the ``right participants'' as part of her broader commitment to research ethics.  P11 ensured that appropriate consenting procedures were in place while conducting research with his participants, using consent as an example of a part in relation to research ethics as a whole. P08 focused on alignment and validity of instruments and the translation of outcomes to inform future design work, revealing interest across multiple stages of engagement in relation to research ethics. In these examples, \textit{research ethics} was used to as a cluster of values or attitudes that linked towards a broader notion of ethical engagement.

\subsubsection{Lens}
Practitioners used synecdoche as a \textit{lens} to represent and illustrate their ethical commitments as a perspective. The lens served as an additional layer that was added by the practitioner to foreground certain aspects (or ``parts'') of their broader ethical ``whole.'' The most prominent lens was stressed through the language of being ``user-focused.'' 
For example, P01 mentioned the ``parts'' that became foregrounded when considering their lens of being ``user-focused,'' including: ``\textit{advocating for the user, making sure things are inclusive and accessible, being conscious of privacy, and surfacing information to the users.}'' This set of parts became related both to a goal of being user-focused and as a means of doing ethical work, as P01 is ``\textit{connecting with users often.}'' She links these elements together by framing her work as advocacy for users, connecting the parts to a broader ethical perspective: ``\textit{I represent users and I am a voice for people who a lot of the time don't have a voice.}'' Other instances included P02, who refers to himself as a ``representative'' of the end user of the app; P09 referred to iteration of his design outputs based on ``user analytics'' as a proxy for user-focused work; and P08 posed questions to critically reflect on his process asking: ``\textit{How do you stop it influencing and not start manipulating the users?}'' In these examples, practitioners focused on the user as a \textit{lens} for languaging their design process, decision making, and evaluation of their products that reinforced a broader conception of what made these activities ethical.
 
\subsection{Metonymy}
Metonymy was used by practitioners to describe ethics by \textit{replacing} ``ethics'' with a related term or concept. Practitioners used metonymy in two ways: 1) \textit{Explicit replacements} where practitioners consciously replaced ethics with terms that suited the practitioner’s context (i.e., ethics as \textit{regulations} by P11), and 2) \textit{Implicit replacements} where practitioners replaced ethics with terms indirectly and we as researchers inferred these replacement from their conversation (i.e., \textit{morality} by P01).

\subsubsection{Explicit Replacements}
Practitioners used metonymy to explicitly and consciously replace the term \textit{ethics,} drawing from personal contexts and current/past ecologies. This type of replacement was easily identifiable due to the participant's explicit description of ethics in relation to their work ecology or project outcomes. As a prominent example of an explicit replacement, P11, a UX Researcher, had a moment of self-awareness towards his use of metonymy during the co-creation session: ``\textit{It's interesting when we don't call this ethics. We call this regulation. Maybe if we all call this ethics, everyone would be more bought into it}.'' In this instance, P11 outwardly recognized the way he explicitly used regulation as a replacement which connected to his work context, where finance and is highly regulated through FCA (Financial Conduct Authority) policies. Further, this replacement through metonymy illustrated how languaging ethics as ``responsibility'' could illuminate certain action possibilities and shut down others; for instance, reducing the number of actors that are responsible for assuring ethical outcomes. In another example, P07 used \textit{compliance} as a replacement for ethics within this particular instance: ``\textit{For example, simple thing could be like, `Oh, we have these policies.' So let's say if we are defining a product, a feature---let's say scanning attachments---should be PCI [(Payment Card Industry)] compliant.}'' While talking about the ethics of a product, P07 also voiced ethics using the language of compliance and amplified it using an example.

\subsubsection{Implicit Replacements}
Practitioners also used metonymy by subconsciously replacing the term \textit{ethics} in conversation, interacting with the co-creation material using terms that indicated their tacit conceptions of ethics. This type of replacement does not have easily identifiable points of origin and required us as researchers to make inferences. For example, P01 replaced the word ethics within their discussion with notions of morality and legality, without ever clearly identifying which term aligned with what they perceived to be ``ethical'': ``\textit{I've talked to other people about, what's legal isn't always necessarily what's moral, which sometimes morality is a little higher, especially since like individuals have a very different moral compass than compliance versus delivery.}'' Although her language of ``legality'' emerged from the context of an activity where she referred to the ethical dilemma of ``Legal vs. Ethical'', P01 still subconsciously replaced the term ethical with ``moral,'' unprompted. P01 continued to use these terms interchangeably, and we inferred that the two terms likely have a similar meaning for this participant. In another example from P05's discussion, the practitioner remarked: ''\textit{as a nonprofit, you want to be altruistic and then just doing what's good for the people.}'' In this case, being \textit{altruistic} and ``doing what's good for the people'' was a replacement for being ethical or engaging in activities that could be deemed to have positive ethical good associated with them. 

Overall, the instances where implicit replacements occurred were much more common than explicit replacements. Metonymy that was languaged through implicit replacements by practitioners revealed their different conceptions and manifestations of ethics through similarly abstract concepts such as ``morals'' and ``values.'' Through metonymy, the embedded and largely tacit structures that related to definitions of ethics became habitual for the practitioner, with conflation or substitution of terms not always knowable through post-hoc analysis. 

\subsection{Antithesis}
Antithesis was used to describe ethics by \textit{contrasting} two or more different ideas relating to ethics, frequently framing these comparisons around what qualifies as being ethical or \textit{not} ethical. Practitioners used antithesis in three ways in relation to the practitioner's own experience (internal) or other example cases (external): 1) \textit{contesting using internal structures}: where practitioners reflected on their mindset and perception of what is or is not ethical based on their own professional experience and past decision-making; 2) \textit{framing using external structures}: where practitioners used examples or cases of existing services to construct what they felt qualifies as being ethical or not; and 3) \textit{extending external structures to frame internal structures}: where practitioners used examples as a point of reference to present what constituted their own ethical responsibilities and boundaries.

\subsubsection{Contesting using internal structures} 
Practitioners used antithesis to contest their own decision making practices in relation to their ecological setting, project decisions, or knowledge relating to ethics. A few practitioners presented decision making that involved dilemmas that they had faced and contested in order to build what they felt were ``ethical'' outcomes. For example, P05 shared her criteria for determining the ethical nature of a project decision, stating: ``\textit{I think selling data privacy would probably be the last route to go. Like if we can find any better revenue model, we wouldn't go that route.}'' She used antithesis to contrast revenue models, revealing how she would prioritize these differing models in ways that oriented one as more ethical and one as less ethical---linking language to potential action and value orientation. Another frequent use of antithesis was when practitioners framed what \textit{does not} require their ethical responsibility. For example, when P08 was a software engineer, he marked the boundaries of his responsibility by saying ``\textit{my only job is to basically write the code that achieves those features, that implements those solutions. So I'm not worried about what are the real life applications, because it's, unless it's too dangerous---as a software engineer, I mean, it's not my responsibility.}'' In this case, P08 drew the boundaries of what constitutes his ethical responsibility: to implement the solutions and to not ``worry'' about real life applications. Another instance of antithesis is when P11 shared a story about his contract team made an agreement with one of their insurance brokers where they were ``\textit{rigging things}''---an ethically problematic instance where ``\textit{they would only offer our competitors worse rates than they'd offer us}.'' He framed this act as ``\textit{not ethical [\ldots] because we weren't doing what's best for the customer.}'' He added that his company was fined for this action as it was not an ethical decision from a regulatory perspective either. Through this internal case, P11 clearly identified what was ``not ethical'' for him by using antithesis as a rhetorical tool. 

\subsubsection{Framing using external structures}
Practitioners used antithesis by providing examples or cases of existing services (external structures) to frame what was ethical or not, according to them. P06, a software engineer, described how designers are ``\textit{intentionally making their design more clickable.}'' She extended this statement by noting: it is ``\textit{good because when people who choose to spend time on this [interaction] are having fun, but then also it's maybe you're letting people get addicted to those stuff.}'' Here, P06 used antithesis to language the potential consequences of design decisions, particularly contrasting outcomes that relate to fun vs. addiction. In another instance, P07 uses the communication platform Slack to describe an ethically problematic hypothetical: `\textit{``[if they] are leaking [user data] and using it elsewhere, that is unethical.}'' Here, he used antithesis with the support of an example of a consumer facing application to define what would be \textit{unethical} for him in terms of misusing user data. P02 frequently used external structures, comparing and contrasting two existing services to language his perception of what product ethics meant to him. He drew on the example of apps ``\textit{which have a very good business proposition to offer, but the design was pathetic and which are not used by user.}'' He specifically contrasted examples using eBay and Amazon and stand-ins, saying: ``\textit{There are lots of products which are cheaply available on eBay, but people used to prefer to shop from Amazon because eBay has the credibility that people think they are fake or counterfeit perception. Whereas, Amazon has this credibility of customer support. Like if you get a bad rug, Amazon will definitely help you out. That becomes the key value for a user mind.}'' In this case, he drew on his ethics-focused differentiation between two existing services in order to language his perception of ethics as being focused on products having ``credibility'' and matching user expectations. 

\subsubsection{Extending external structures to frame internal structures}
Finally, practitioners used antithesis by providing examples or cases of existing services (external structures) to reflect on their own ethical boundaries, responsibilities, and awareness (internal structures). In this type of antithesis, the external structures acted as a reference point for practitioners to reflect on how they could not do something similar in their organization, or what would be outside of their own ethical boundaries. For instance, P11 mentioned Google ads as an example, saying: ``\textit{it's all good for Google just to stick some advertised links up at the top. So, all search results aren't equal. The ones at the top of the Google get paid for it.}'' He used this case as a reference point to contrast his own decision-making capacity: ``\textit{We can't do that. You know! We get fined a lot of money cause we're not doing what's best for the customer. The customers have to find the best price and if we're putting things that aren't the good price at the top.}'' Here, he contrasted the lack of user-centeredness of Google ads with the design decisions he is able to make which is aligned to what regulatory requirements mandate is ``best'' for the user. While discussing the responsibilities of different professional roles, P07, a designer, mentioned ``[convincing] PMs that if you do achieve this kind of business objective, it should not be at the cost of the person.'' Here, he used antithesis by using the PM's professional role as a reference point, evaluating that achieving business goals would be considered a part of a PM's role while also drawing a boundary to contrast that PM goal with business goals that ``should not'' cost the user and negatively impact their experience with the service.  

\subsection{Amplification}
Amplification was used to describe ethics by sharing detailed descriptions or examples that \textit{illustrate} practitioners' ethical awareness, responsibility, commitment, or action. Practitioners frequently used amplification, drawing from or leading towards synecdoche, metonymy, or antithesis tropes which provided them a foundational language through which to convey their conception of ethics. Practitioners used amplification in three ways: 1) \textit{Augmenting through description} where practitioners provided additional details or examples to support their description of ethics; 2) \textit{Illustrating through internal structures} where practitioners shared their personal or professional experiences (internal structure) to provide details of their ethical stance; and 3) \textit{Illustrating through external structures} where practitioners provided a case (external to the practitioner) to describe their perception of ethics. 

\subsubsection{Augmenting through description:} Practitioners employed amplification by providing a description and explicitly defining their ethical commitment, values, or responsibilities. Using synecdoche as a rhetorical tool, practitioners established various values or elements of their ethical stance they focused on, such as bias (P03), sustainability (P05), user representation (P01), being ``civic-minded'' (P04), or ``PCI compliant''(P07). These terms were often vague upon first use in conversation, but practitioners extended these terms through amplification to define what these terms meant for them. For example, P03 described various kinds of bias that might exist in the data. She extended talking about bias by saying ``there might be some contextual bias in the data itself,'' using amplification to more narrowly frame her meaning of contextual bias, described as follows: ``\textit{So essentially programmed elements of the algorithms are, you know, fairly accountable for certain kinds of contexts.}'' P05 related their ethical standpoint to being civic-minded, which she describes as her desire ``\textit{to be a responsible citizen and be responsible person and do good in the world.}'' This form of amplification often provided further clarity into how part-whole relationships introduced through synecdoche impacted the practitioner's ethical stance towards a product or society. 

\subsubsection{Illustrating Internal Structures:} Practitioners used amplification by sharing examples from their personal or professional experience (internal structures) that supported their ethical stance. P07 described an instance in their collaborative work where there was concern about ``taking credit of someone else’s work. Working out with a team, there is certain ethical consideration you need to be taking while collaborating as well.'' Here, he drew on his everyday collaboration with his colleagues and what aspects of false crediting would constitute unethical behavior. P12 also mentioned that they seek to not confuse the users by providing adequate information that will help them to use the platform meaningfully, accomplished through articulating clear and explicit design decisions. She specifically stated that her goal was ``\textit{always to make sure that users are constantly aware of what they are agreeing on. So we try to not make anything and put in place it and to state everything explicitly.}'' 

\subsubsection{Illustrating through external structures:} Practitioners used amplification by providing cases of existing applications (external structures) to elaborate their perception of ethics. For example, P01 used Uber and Lyft in California as examples to clarify their concerns about privacy concerns in ubiquitous digital products and how these concerns could relate to various policy requirements. She described this amplification through examples as follows: ``\textit{for example, privacy wasn't a big concern before there was the rise of ubiquitous computing. With that brought a lot more privacy concerns and that affects the policies that the government faced. Like with Uber and Lyft in California, they helped author legislation that benefits them.}'' P02 used references of a range of existing digital services such as Facebook, Amazon, eBay, Netflix, and Paytm to compare and describe how these services employ user-centric solutions to improve their business offers and value: ``\textit{in terms of revenue, let me call this number and the end of December of 2020, Paytm was doing the transaction or worth rupees 5 billion. Whereas the second nearby competitor PhonePe was doing a translation of around 2.1 billion. You can see the 2x difference.}'' He used this business case data to describe the adaptability of the apps that relied upon habit-forming features and how these cases guided him when he made decisions as a product manager that would improve the KPIs in his organization. In this latter case, the amplification drew attention away from the source of ethical engagement and towards other metrics of success (e.g., KPIs).

\section{Negotiation of Ethical Standpoints through Languaging} \label{cases}
In this section, we answer RQ\#2 by visualizing and describing how three participants used language to articulate their ethical standpoint in relation to their ethical ``core,'' internal emergence, ecological emergence, and ultimate boundaries. Each case illustrates the individual and ecological complexity of the ethical standpoint of one technology practitioner, using infrastructural and direct vignettes to demonstrate how rhythms of languaging formed and solidified the practitioner's conception of ethics. 

\subsection{\textit{There is always a trade-off when it comes to ethics}}
\begin{figure}
    \centering
    \includegraphics[width=\textwidth]{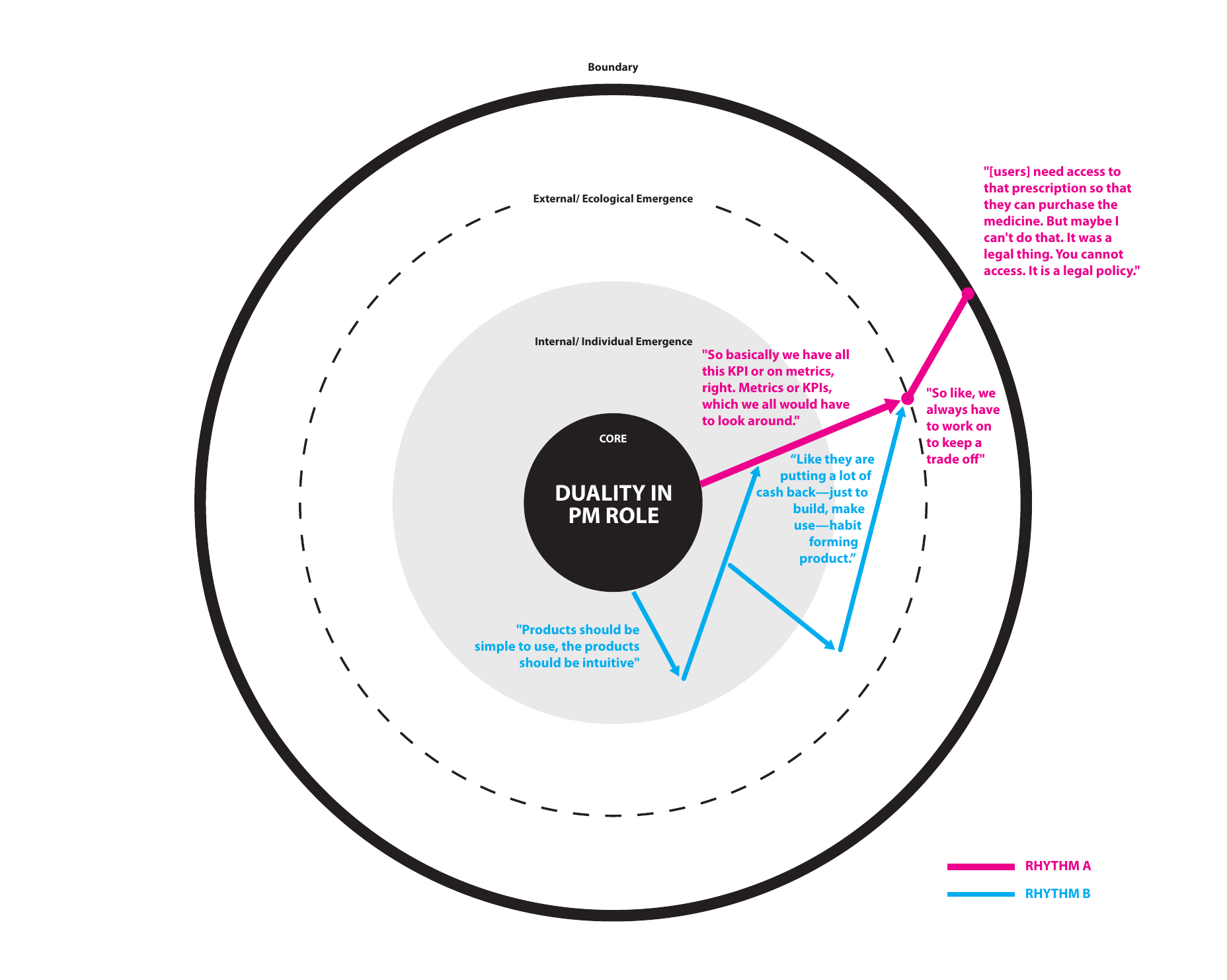}
    \caption{Layers of Languaging of P02}
    \label{fig:1P2case}
\end{figure}

P02 is a product manager in Enterprise (B2B) environment with around five years of industry experience. He currently creates healthcare digital services with his end users primarily consisting of doctors. The stated responsibility in his professional role includes receiving ``goals or instructions'' from the company CEO and translating them into a ``plan of action'' for his team of software engineers, designers, and data scientists. As shown in Figure \ref{fig:1P2case}, P02's core is drawn from his professional role of being a product manager where he has to balance both business goals and user goals. This balancing leads to a duality in his core, informing two prominent rhythms of languaging that summarize his discourse and potential for action concerning ethical matters: Rhythm A, which describes his goals towards increasing KPIs and number of transactions to reach business goals; and Rhythm B, which bounces off rhythm A as he uses the language of being \textit{user-centric} to seek to frame and reach those business goals. 

As shown in Figure~\ref{fig:1P2case}, his professional responsibility frames his \textbf{core} which consists of a duality where on ``\textit{an organization level, I am focusing more toward achieving that particular goal, which I got from my CEO}'' and on ``\textit{a personal level, I had to detach myself from that particular goal [...] I am thinking myself as the representative of the end user of that particular app.}'' To negotiate this core, different rhythms of languaging illustrate how this participant engages with multiple priorities, primarily drawing from \textit{business-oriented strategies} such as increasing KPIs, number of transactions, and other business metrics (rhythm A). He supports business-oriented language in his work by framing this language through \textit{user-focused goals}, represented by an interlaced Rhythm B. For instance, metrics were related in language to designing a simple UI, ensuring that interfaces are intuitive, and considering the usability of digital products. The manifestation of metrics into product features represent the \textbf{internal emergence} of his core, where he uses synecdoche as a means of languaging his approach to ethics. To further describe how he sought to support user-focused goals while maintaining his ultimate aim of increasing numbers of transactions, he used the concept of \textit{adoption} of these digital services by the users through a description of particular platforms through which to acknowledge his ethical positioning. For example, he uses antithesis as he compares Facebook and WhatsApp to talk about the simple intuitive user interface design of WhatsApp which allows ``\textit{adoption or acceptance in villages than Facebook. So Facebook has a very high fidelity detail on the design. But whereas in WhatsApp, the design is very simple, that’s why WhatsApp is more acceptable in villages, in cities, in competitor than adoption of Facebook.}'' 

As his languaging extends into his \textbf{ecological emergence}, his user-centric language around business-focused goals as a duality is illustrated through particular kinds of design decisions that relate to the adoption and adaptability of digital products. Three key kinds of ecological emergence reflected a commitment that was focused more on his business commitments than his commitments to users. First, he referenced toying with user emotions, concluding: ``\textit{We have to play with emotion or yeah. But one thing is for transaction, like to making you do the solution, I will have to play on the emotional part.}'' Second, he described how he engaged in building habit-forming products to retain users. And third, he leveraged features that are a priority to the users, for example, ``\textit{So doctors, they care a lot about of privacy, right? Like for every product, which I market to them, privacy is a very important key area, so we have to market about them.}'' 

As the two key rhythms extend from individual emergence to ecological emergence, his relational ethical boundaries demonstrate how his core commitments result in accepting that his decision making ``\textit{always [results in] keep[ing] a trade off}.'' The trade-offs exist when he, by virtue of his role, has to consider what his boss or the CEO instructs as the business goals and his role as a professional tends to abide by those ``instructions.'' His perspective towards the trade-offs is a manifestation of the duality of his role and responsibility which brings the two rhythms of his justification together at the permeable boundary of \textbf{ecological emergence}. Although he accepts that there are always trade-offs in ethical decision making, P02 elaborates upon the line he is not willing to cross: \textit{legal policy}, as defined by his ecology, represented as an ethical \textbf{boundary}. For instance, when a patient attempted to purchase their medication directly from his company which only was available through the doctors, he reflected, ''\textit{\ldots it was a legal thing. You cannot access. I felt bad, but I can't help that particular thing.}'' He is unwilling to cross the boundaries of legal policies as a practitioner, even when it might benefit the end user, demarcating his ethical standpoint. 

\subsection{\textit{This is/is not my responsibility}}

\begin{figure}
    \centering
    \includegraphics[width=\textwidth]{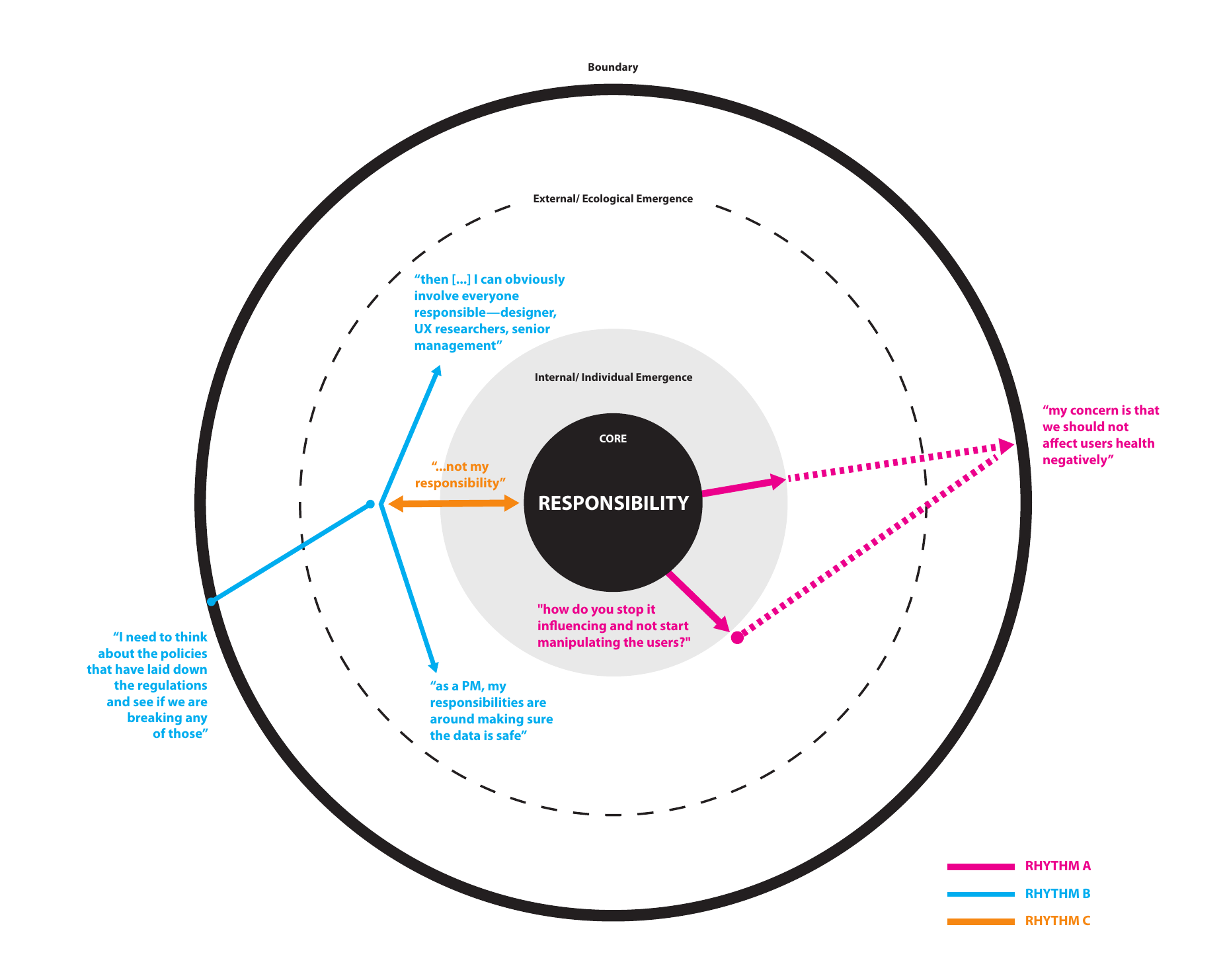}
    \caption{Layers of Languaging of P08}
    \label{fig:3P2case}
\end{figure}

P08 is a product manager working in an Enterprise (B2B) environment who has previously been a software engineer with more than 8 years of industry experience. As shown in Figure \ref{fig:3P2case}, P08's core is formed based on his languaging of ethics in relation to personal and professional ``responsibility.'' He represented his core in three distinct rhythms in the schema: Rhythm A, which defines what he languaged as his personal responsibility to ensure so that the user is not harmed, establishing his ultimate ethical boundary; and Rhythm B which builds on his professional responsibility regarding what his role is responsible for and how the responsibility is distributed, with an embedded Rhythm C that defines how his different roles have indicated what he is \textit{not} responsible for. 

In rhythm A, P08 defines his \textbf{boundary} of not harming the users as follows: ``\textit{my concern is that we should not affect user's health negatively}'' in reference to creating user engagement within an app. This boundary is supported by his core of responsibility mediated through his professional responsibility as the designer of the technological product not to harm users and his internal sense of responsibility that seeks to differentiate influence from manipulation. This tension between ``\textit{how do [I] stop it influencing and not start manipulating the users}'' and not affecting user health negatively is negotiated in the ecological emergence layer through product-focused questions such as ``\textit{So if we have the product as a game for kids. So how do we limit users screen time?}'' and ``\textit{the product manager responsibility would be to identify ways to stop users from playing, let's say two hours or three hours}.''

In rhythm B, P08 frames and defines his core---responsibility---as what he must do in his professional role of being a product manager, which he consistently voiced as follows: ``\textit{my responsibilities are around making sure the data is safe.}'' This responsibility is defined within the ecological emergence space in relation to his professional role, and in relation to others that he may need to delegate tasks to in order to achieve his goal: ``\textit{then [\ldots] I can obviously involve everyone responsible—designer, UX researchers, senior management.}'' These responsibilities are then also related to boundaries he is unwilling to cross, such as existing regulations: ``\textit{I need to think about the policies that have laid down the regulations and see if we are breaking any of those.}'' Because P08 was in the process of shifting from a software engineer to a product manager, an embedded rhythm C further describes what he felt was \textit{not} his responsibility in his ecology. As a product manager, the responsibility of his role was to consider regulation and policies, but when he was a software engineer he says regulatory consideration ``\textit{was not my problem}'' because his professional responsibility didn't allow him to think about it. As part of his new role as a product manager, he consistently languaged the responsibilities of other professional roles saying: ``\textit{The designer [\ldots] their responsibility is to implement the UI/UX aspects of the product}`` and ``\textit{the role of the data scientist is to measure the product.}'' This ``othering'' of responsibility in P08's ecological emergence allowed him to frame and shape his own PM responsibilities but also puts constraints on his own responsibility that shaped his core.

\subsection{\textit{We do not call it ethics, we call it regulation}}

\begin{figure}
    \centering
    \includegraphics[width=\textwidth]{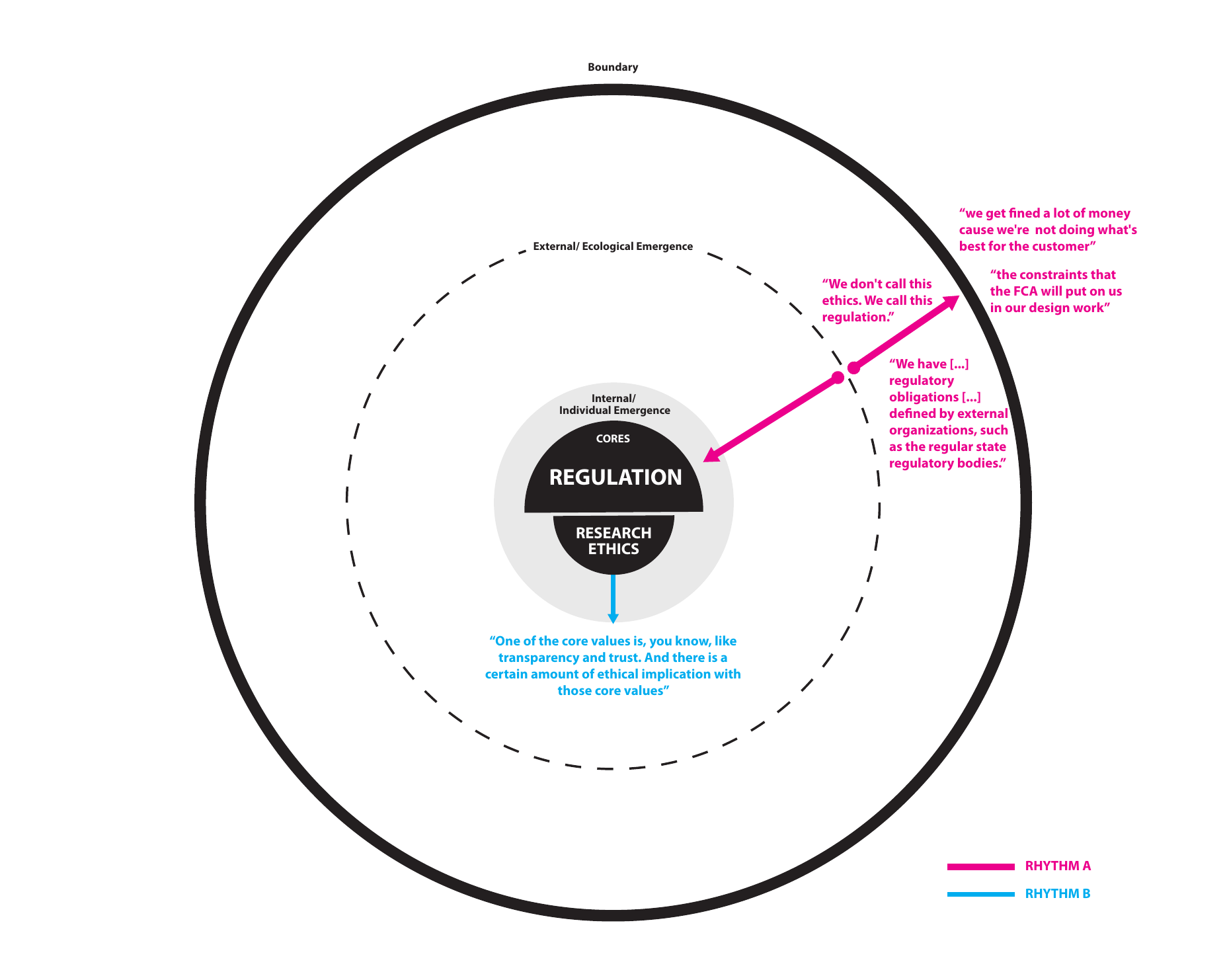}
    \caption{Layers of Languaging of P11}
    \label{fig:4P2case}
\end{figure}

P11 is a user-experience (UX) researcher in an Enterprise (B2B) environment with four years of industry experience. Working in a financial technology company (B2B), his professional responsibilities as a UX researcher entail researching and developing frameworks to make his company's financial product more usable and user-friendly for a team of legal advocates, product managers, designers, and software engineers. He also must ensure that their product conforms to the financial regulatory requirements defined by the Financial Conduct Authority (FCA) in Britain. As shown in Figure \ref{fig:4P2case}, P11 has two main cores, one stemming from his ecology which focuses on ``regulation'' and the other---research ethics---from his professional role of being a researcher. The two cores led to two rhythms of languaging: 1) Rhythm A, where regulation forms the \textbf{core} of P11's language of ethics drawn from ecological emergence of him working in the organization which is highly regulated by FCA guidelines; and 2) Rhythm B, where research ethics forms the \textbf{core} drawn from his role as a UX Researcher which includes a consistent reference to informed consent, research subject protection, and improving trust and transparency with participants. Through his languaging, P11 decisively engaged with notions of either being ethical or following regulations that were heavily drawn from his ecological setting. In comparison to the prior two cases, this engagement enlarged the ecological emergence space as he languaged his ethical role.  as compared to the two cases detailed above. 

In rhythm A, P11's languaging is drawn from his \textbf{ecological emergence} directing towards framing his core of ``regulation'' as: ``\textit{we have ethical regulation, regulatory obligations in terms of what we do now, who defines those it's in part defined by external organizations, such as the regular state regulatory bodies.}'' Given the framing of ethics embedded in his organization, he refers to ethics as regulation using metonymy; his quote: ``\textit{It's interesting when we don't call this ethics. We call this regulation. Maybe if we all call this ethics, everyone would be more bought into it.}'' His language has been largely influenced by regulatory constraints defined in the ecological emergence layer. For example, he shares as example of ``\textit{Google ads---it's all good for Google just to stick some advertised links up at the top. All search results aren't equal, the ones at the top of the Google get paid for it. They go to the top, we can't do that.}'' He uses amplification to illustrate what kinds of design decisions form his ecological emergence. For P11, this ecological emergence also extends to form his ultimate \textbf{boundary} formed by regulation because not following the FCA guidelines will result in their organization getting ``\textit{fine[d] because we weren't doing what's best for the customer.}'' This rhythm forms the primary languaging of P11's ethics which are constrained, translated, and defined by regulation. 

In rhythm B, P11's language stems from the core of research ethics which is drawn from his educational background and professional role as he strives to adhere to all the current best practices like issues of trust and transparency while conducting research. As part of his individual emergence, P11 articulates core values that are different from his regulatory core, as he reflects on his personal and professional role as an individual: \textit{``One of the core values is, you know, like transparency and trust. And there is a certain amount of ethical implication with those core values.''} This individual emergence of ethical standpoint using the framing of research ethics is then extended to the ecological emergence layer, where P11 considers how these values impact product decisions: \textit{``We need to make sure that this is compliant to current best practice and accessible.''} 

\section{Discussion}
In this paper, we have illustrated how technology practitioners use language to reveal their conceptions of ethics through a range of rhetorical tropes, creating a space for potential action across individual and ecological layers. In the following sections, we will describe the potential uptakes of both the set of rhetorical tropes and layers of languaging schema in supporting ethics-focused engagement across a range of settings. First, we synthesize how the use of various rhetorical tropes led participants to flatten complexity, often resulting in a relatively reductionist view of technology ethics. Second, we describe how movements illustrated through our schema allow for a fuller accounting of personal and ecological influence as practitioners language ethics, considering how the expansion and contraction of ethics as a conceptual space can be supported through languaging. 

\subsection{Rhetorical Reductionism}
The rhetorical tropes we have described in our findings---synecdoche, metonymy, antithesis, and amplification---are common forms of languaging that aid us in constructing our social world, revealing opportunities for action. Building on our findings in Section~\ref{tropes}, we seek to identify what these various rhetorical tools tell us about practitioners' framing and conceptualization of ethics, revealing how language brings ethics to life as a construct in their everyday work. We summarize how the use of each of these rhetorical tropes aided practitioners in describing their ethical standpoint, while also frequently reducing complexity in ways that made these ethical standpoints difficult to inspect, challenge, or ensure mutual understanding. 

\begin{itemize}
    \item \textit{Synecdoche} allowed practitioners to present their ethical positioning in the form of a part--whole relationship, giving a linguistic \textbf{focus} to their conception of ethics that was driven by what we might call ``part-ethics.'' These partial descriptions of ethics pointed towards an often implicit ethics whole, drawn from their individual experiences, organizational context, disciplinary rhetoric, project domain, or societal outlook. However, in most cases, the linguistic focus facilitated through synecdoche drew important, yet often not fully considered, boundaries through which practitioners could engage. By using part-ethics, a certain aspect of ethics such as accessibility or privacy stood in for the whole of ethics, potentially \textit{restricting} the practitioners ability to later expand their rhetoric of ethics beyond only the piece that originated the languaging. Synecdoche frequently acted as the foundation of practitioner's language, facilitating precision in language while also having the impact of constituting the whole of ethics as ephemeral and ill-defined. Upon establishing a linguistic frame using synecdoche, the use of other rhetorical tools typically trickled down from the part-ethics, further streamlining practitioner's rhetoric around ethics but perhaps resisting an exploration of the assumed intersubjective space around the broader conception of ethics as a whole.  

    \item \textit{Metonymy} allowed practitioners to replace the term ``ethics'', with the impact of either \textbf{restricting or expanding} their rhetoric based on the semantic and epistemological assumptions embedded in the terms used. For instance, practitioners using the term ``regulation'' substituted a moral philosophy epistemology for a legal epistemology, shifting perspectives regarding what kinds of knowledge is valued, what kinds of rationale are acceptable, and which frames of amelioration might exist. In contrast, by using the language of morality or values, the epistemological assumptions in the replaced term aligned with the broader conception of ethics, but exchanged one type of ambiguity for another. The semantics of the replacement terms used allowed the practitioners to re-focus their languaging of their ethical responsibility, but this refocusing came at the expense of either shifting complexity into another epistemological sphere (which could have the added complication of reducing the wrong \textit{sorts} of complexity) or maintaining ambiguity while assuming a shared intersubjective understanding of the broader concept of ethics. 
    
    \item \textit{Antithesis} allowed practitioners to be \textbf{definitive} in their languaging regarding what they felt was ethical or not ethical through the use of linguistic contrast(s). Antithesis was effectively used by practitioners to establish a linguistic ``gestalt,'' contrasting appropriate ethical behaviors and standpoints with ultimate boundaries they would be unwilling to cross. However, these binaries introduced through antithesis were also impacted by earlier definitions of ethics framed through synecdoche, which allowed for internal clarification of what was or was not ethical, but did not allow for broader questioning of how the part-ethics related to ethics as a whole. For instance, practitioners talked about accessibility as their core focus of ethics. Using antithesis, they were only able to build on what they considered to be ethical or unethical within the boundaries of accessibility rather than expanding the conversation to explore framings of ethics using accessibility as a lens, or by contrasting accessibility with another framing of ethics.     
    
    \item \textit{Amplification} allowed practitioners to \textbf{contextualize} their understanding of ethics by using specific examples, professional experiences, or case studies. Practitioners either used amplification to talk about a specific focus through an example, later leading towards Synecdoche, Metonymy, or Antithesis, or began with an existing linguistic focus that they then amplified through the use of an example. Through the examples provided, practitioners primarily used amplification to solidify their introduction of part-ethics (synecdoche), replaced terms (metonymy), or boundaries (antithesis). Due to limitations of this trope, amplification could not be used to directly challenge the inherent limitations of these other framings of ethics, and in many cases, the examples themselves implied certain ethical standpoints without allowing them to be more directly inspected or evaluated.  
\end{itemize}

As we have illustrated, practitioners' languaging of ethics often resulted in a \textit{reductionist} perspective that was difficult to escape---in part due to the use of concepts or terms to frame, rather than problematize, their ethical engagement. Practitioner languaging using these common rhetorical tropes tended to be restricting, confining, or definitive rather than expanding, generating, contesting, or evolving---assuming a shared understanding of ethics as a broader category of knowledge, inquiry, or guidance for action. What is the impact of this reductionism, and why does language matter? First, we observed that this reductionism through languaging often shut down inspection or contesting of potential dimensions of ethical awareness, knowledge, and action. This lack of inspectability or ability to objectivate---or bring tacit assumptions into explicit language---shows that the language being used to express complex concepts such as ethics has flattening characteristics which discourages or resists challenge. However, these same tropes could be used to problematize one's ethical standpoint, identifying tensions in different ethical ``cores,'' calling out edge cases which are well addressed by one framing of ethics but completely neglected in another, or recognizing that one's ethical standpoint has been problematically flattened. Second, their appears to be an assumption that practitioners are creating a meaningful and intersubjective space where mutual understanding occurs---both in research studies like the one we report on here, but also in authentic practice contexts where practitioners are conversing with each other about ethics. Without objectivation, practitioners may assume that their underlying subjective expectations embedded into their language are shared, but as we have illustrated, many common rhetorical tropes can tend to ambiguate rather than disambiguate the meaning of higher order concepts if they are not challenged and contrasted; thus, practitioners are likely assuming different intersubjective spaces that may be actively in tension, yet with no ability to reveal the differences in language. Engaging in objectivation through simple questions such as ``What do you mean by ethical?'' or ``Is accessibility (or a similar stand-in for ethics) the only thing we should be focusing on?'' may open up the rhetorical space for questioning, revealing key deficiencies or differences in perspective across a set of technology practitioners.

\subsection{Ecological Influence in Languaging Structures} \label{connectiontoliterature}
We have previously described how individual technology practitioners' conceptions of ethics and potential for performance exists within an ecology. In this section we seek to more fully describe the ecological influence on practitioners' languaging structures. Building on the cases and schema used to illustrate the cases in Section~\ref{cases}, these ecological elements aid us in further describing what infrastructural concerns may relate to the emergence and conception of ethics in technology practice, and allow us to build further connections with moral philosophy, HCI, and STS scholarship that has the potential to inform future work. As we illustrate the co-relations that emerged in our \textit{Layers of Languaging} schema and the three cases, we aim to highlight the complexity in languaging ethics in ways that correspond to the felt complexity of technology practice: 

\begin{itemize}
    \item \textbf{\textit{Core} <--> Activation of/Alignment with Ethical Paradigms} In our schema, the ``core'' indicates the frame or guiding metaphor for the practitioner's conception of ethics that encapsulates how this conception emerges in language. We observed that this metaphor is aligned with one or more ethical theories as defined in moral philosophy, as shown in the three evident cores: a duality of professional role, responsibility, and regulation + research ethics. For instance, P02's core on the \textit{duality of his role} took a deontological ethical stance where he is following his ``duty'' of being a product manager as it is framed as a job role and/or discipline, with the balancing of the duality relating to a pragmatist ethical stance when the roles were in conflict to guide the ``tradeoff.'' P08's core of what constitutes or does not constitute his \textit{responsibility} and the impacts of that responsibility took on a virtue and consequentialist ethical stance, frequently aligning with notions of professional ethics. Finally,  P11's core of \textit{regulation} took a deontological ethical stance by following the practices to abide by regulatory requirements and a consequentialist ethical stance by seeking to anticipate the impacts of failing to follow those regulations on the product, the company, and himself. Importantly, these metaphors appear to align in certain ways with these existing ethical paradigms, but the ways that practitioners languaged ethics were not constricted, constrained, or even framed by these ethical paradigms, but rather overlapped and interacted in ways that emerged naturally in language.
    
    \item \textbf{\textit{Internal/Individual Emergence} <--> Co-constitution of various Identity Claims} The process of internal/individual emergence illustrates the activation of the core as it was considered and structured by the practitioner in relation to both their professional role and person. This individual emergence points towards identity claims that they considered and then later operationalized as part of an ongoing co-constitutive process. Within this process, we can relate the co-constitution to Chivukula et al.'s~\cite{chivukula2021identity} vocabulary of identity claims, where many identity claims can be considered ``true'' for a person, but certain identity claims are particularly foregrounded or prioritized to create a sense of coherence or connection. In our analysis in this paper, we see a co-constitution of individual emergence resonating with this previous vocabulary, expressing a relational set of claims which are described ``in relation to their role (as a technology practitioner), the social elements of that role (team or organizational), and their understanding of self (a manifestation of their constructed identity)''~\cite{chivukula2021identity}. In this way, the formation of the individual emergence is co-constituted in relation both to the guiding ``core'' metaphor and to the potential for professional and ecological engagement. This formative space serves as the incubator for an individual to internally build out their ethical perspective and consider which values or identity claims they wish to privilege. 

    \item \textbf{\textit{External/ Ecological Emergence} <--> Alignment with Ethical Mediation, Value Levers, and Soft Resistance} Ecological emergence shapes the languaging of practitioners based on practices that are required, prescribed, or constructed in relation to their ecological setting. The language that emerges in relation to ecology builds upon the core and internal emergence, but is not restricted or confined by these elements of a practitioner's ethical standpoint. In contrast, some elements of one's core or internal emergence are shaped by external ecological forces or emerge and then later shape the broader practice ecology (cf., fulfillment or nullification or in \cite{Watkins2020-zr}; creating spaces and visibility for values through soft resistance in \cite{Wong2021-pv}; value levers to open up conversations that might impact practices in~\cite{shilton2013values}). The movements and rhythms within the ecological space can also be explored through Gray and Chivukula's notion of \textit{ethical design complexity} \cite{gray2019ethical}, where the practitioner continuously mediates their ethical standpoint across individual values, organizational practices, and applied ethics. In our schema, this ethical design complexity is demonstrated as a dynamic and ongoing process, through which a broader conception of the ``whole'' of ethics is languaged through a range of interactions among elements of the ecological setting in ways that cross multiple layers and can result in various levels of fluidity or solidity. For instance, P11's languaging of ``ethics is regulation'' was primarily drawn from his ecological setting, a financial service company which is highly regulated. In this case, ecological emergence dramatically influenced P11's languaging where his core, internal emergence, and ultimate boundary all came to be defined by ``regulation.'' In contrast, P08 negotiated their ethical design complexity through ecological engagement as part of their professional role, while also engaging values that related to their responsibility as an interaction between their core, internal emergence, and ultimate boundary.
    
    \item \textbf{\textit{Boundary} ---> Result of Ethical Paradigms, Identity Claims, and Ethical Mediation} An ultimate boundary is formed by practitioners as they consciously negotiate and objectivate a mix of internal and external emergence elements. This boundary operates as a limit case, but the ethical standpoint that informs this limit case might come from an individual's own personal values, from elements of their professional role or organizational type, or from hypothetical ethical dilemmas that express their ``whole'' of ethics in concrete form. The pairing of core and ultimate boundary express the relative coherence of the ethical standpoint, and also allow for this standpoint to be characterized in relation to one or more ethical paradigms. This pairing also foregrounds particular aspects of the overall set of languaging layers, revealing tensions or alignment. For instance, P08's boundary is built on his internal emergence of considering that his ethical boundary must not effect the ``user's health negatively,'' revealing that their sense of core ``responsibility'' is likely to exist primarily in relation to wellbeing and the potential for harm. In contrast, P11's boundary---regulation---is clearly solidified by ecological emergence based on the type of company he works for, causing regulation to potentially suppress his other core of research ethics in places where tension might arise; for P11, there are also no boundaries that relate to research ethics, positioning this core only as an internal commitment, but one that is unlikely to be concretized as part of a broader ethical standpoint. These languaging relationships demonstrate bidirectional relations, shaping how boundaries influence and frame both internal emergence through prioritization of identity claims and the mediation of these ethical concerns in an ecological sense through elements of one's professional role and organizational structure.
\end{itemize}

\section{Implications and Future Work}
Our findings point towards a range of potential implications that could impact HCI and STS researchers, technology practitioners, and educators. As framed by our discussion, the notion of languaging opens up the space of ethical engagement for fuller inspection in ways that exceed only labeling behaviors as ``ethical'' or ``unethical,'' positioning ethics as primarily about following a code of ethics, or describing ethical reasoning through the lens of one or more ethical theories. The variety of rhetorical strategies that our participants used to language ethics reveals the potential for more precise research engagement in subjective and situated experiences of technology practitioners in relation to their ethical standpoint, potentially joining ethnographic, observational, and interview research that has described challenges from an ecological perspective (e.g.,~\cite{shilton2013values,Wong2021-pv,Steen2015-qw,gray2019ethical,lindberg2020cultivating}), interventionist work that encourages and challenges practitioners to make ethically-motivated decisions (e.g.,~\cite{Van_Wynsberghe2014-wf,Ballard2019-lu}) and impact-driven work that seeks to encourage ethical action through awareness and reliance on moral philosophy (e.g., \cite{Friedman2019-zg,Flanagan2014-hf}). From each of these perspectives, analytic toolkits may be sharpened by attending more closely to both the words that practitioners use to describe their ethical commitments and engagement, and the relative coherence of this ethical standpoint that might impact their ability to act. Both the set of rhetorical tropes and Layers of Languaging schema might be used to better understand the subjective position and internal coherence of a practitioner's ethical standpoint, both as a reflective and analytic tool. Future work could include both analysis of interview, observational, or social media data relating to technology practice to describe what systems of ethics are being languaged and elicitation-focused work where the dialogic properties of ethics languaging can be foregrounded and used as a tool to encourage mutual understanding with a more precise vocabulary. 

Building on our discussion, we also find potential value in considering the ways in which languaging can both illuminate a broader ethical standpoint and also shut off that standpoint from being questioned. While the practitioners we engaged with primarily used rhetorical tropes to reduce complexity (likely a limitation of our co-creation approach, where we did not actively aid them in contesting their worldview or ethical standpoint), we have also identified opportunities to activate language in service of objectivating practitioners' ethical standpoints---bringing tacit assumptions of the whole of ethics into focus and encouraging explicit attention towards the functioning of that standpoint at various layers of languaging. This objectivation work has the potential to be useful, not only for researchers seeking to more fully describe the ethical standpoint of their participants on their own terms, but also to aid practitioners and educators in challenging their own and others' perspective relating to ethics by using the schema and rhetorical tropes as a framework for discussion. As an example, future work could equip educators to encourage technology and design students to form and then question their own ethical standpoint---building awareness of different infrastructural levels of ethical engagement with varying levels of coherence rather than focusing curricula primarily on codes of ethics or ethical reasoning in the abstract.

\section{Conclusion}
In this paper, we have described and visualized how technology practitioners used rhetorical tropes to reveal their conceptions of ethics. Through the analytic frame of languaging, we identified how practitioners used the rhetorical tropes of synecdoche, metonymy, antithesis, and amplification to construct their conceptions of ethics in language, and then visualized the cases of three practitioners to illustrate how ethics is infrastructurally related to core commitments, internal emergence, ecological emergence, and ultimate ethical boundaries. Building on these findings, we identify both the limitations of articulating ethical standpoints through language and ways that language might be better leveraged to inspect, evaluate, contest, and objectivate ethics in order to encourage mutual understanding and attention to the limitations of any one ethical frame. We conclude with opportunities for practitioners, researchers, and educators to use this vocabulary to build awareness and alignment in relation to ethical concerns, while also recognizing the diverse linguistic framings of ethics that can problematize, contest, discourage questioning, or envision new modes of technology ethics. 

\begin{acks}
This work is funded in part by the National Science Foundation under Grant No. 1909714.
\end{acks}

\bibliographystyle{ACM-Reference-Format}
\bibliography{referenceslang.bib}

\end{document}